\documentclass[useAMS]{mn2e}
\usepackage{graphicx}
\begin{document}
\def\simlt{\mathrel{\rlap{\lower 3pt\hbox{$\sim$}}\raise 2.0pt\hbox{$<$}}}
\def\simgt{\mathrel{\rlap{\lower 3pt\hbox{$\sim$}} \raise 2.0pt\hbox{$>$}}}
\def\Msun{M_{\odot}}
\def\Zsun{Z_{\odot}}

\title[Simulating high-redshift galaxies]{Simulating high-redshift galaxies}
\author[Salvaterra, Ferrara \& Dayal]{Ruben Salvaterra$^1$, Andrea Ferrara$^2$ \& Pratika Dayal$^3$\\
$^{1}$ Dipartimento di Fisica e Matematica, Universit\'a dell'Insubria, Via Valleggio 7, 22100 Como, Italy\\
$^{2}$ Scuola Normale Superiore, Piazza dei Cavalieri 11, 56126 Pisa, Italy \\
$^{3}$ SISSA/International School for Advanced Studies, Via Beirut 2-4 Trieste, Italy, 34014}

\maketitle \vspace {7cm}

\begin{abstract}
Recent observations have gathered a considerable sample of high redshift galaxy candidates and determined the evolution of their luminosity function (LF). To interpret these findings, we use cosmological SPH simulations including, in addition to standard physical processes, a detailed treatment of the Pop III$-$Pop II transition in early objects. The simulated high-$z$ galaxies match remarkably well the amplitude and slope of the observed LF in the redshift range $5 < z < 10$. The LF shifts towards fainter luminosities with increasing redshift, while its faint-end slope keeps an almost constant value, $\alpha \approx -2$. The stellar populations of high-$z$ galaxies have ages of 100-300 (40-130) Myr at $z=5$ ($z=7-8$), implying an early ($z>9.4$) start of their star formation activity; the specific star formation rate is almost independent of galactic stellar mass. These objects are enriched rapidly with metals and galaxies identified by HST/WFC3 ($M_{UV}<-18$) show metallicities $ \approx 0.1 \Zsun$ even at $z=7-8$. Most of the simulated galaxies at $z\approx 7$ (noticeably the smallest ones) are virtually dust-free, and none of them has an extinction larger than $E(B-V) = 0.01$. The bulk (50\%) of the ionizing photons is produced by objects populating the faint-end of the LF ($M_{UV} < -16$), which JWST will resolve up to $z=7.3$. PopIII stars continue to form essentially at all redshifts; however, at $z=6$ ($z=10$) the contribution of Pop III stars to the total galactic luminosity is always less than 5\% for $M_{UV}<-17$ ($M_{UV}<-16$). The typical high-$z$ galaxies closely resemble  the GRB host galaxy population observed at lower redshifts, strongly encouraging the use of GRBs to detect the first galaxies.  
\end{abstract}


\section{Introduction}
\label{introduction}
The search for the most distant galaxies, located at the beginning of the cosmic dawn, is
now entering its maturity. The last few years have witnessed a tremendous increase in the data available, and the number of candidates at redshifts as high as $z=10$, corresponding to only half a billion years after the Big Bang. This has been made possible by a combination of new technologies and refined selection methods. In the first class of triggers, it is easy to acknowledge the role of the Hubble Space Telescope (HST). Thanks to dedicated surveys including the Hubble Ultra Deep Field and its predecessor, the Hubble Deep Field, we have been able to collect information on the luminosities and number counts of galaxies located at the end of the reionization epoch. Immediately after, follow-up experiments performed with the newly installed Wide Field Camera (WFC3), yielding sky images in the F105W (Y-band, $1.05 \mu$m), F125W (J-band, $1.25 \mu$m) and F160W bands (H-band, $1.60 \mu$m), have allowed to push the exploration to very faint (e.g. AB mag = 28.8 in the above bands) galaxies as remote as $z=10$. In addition, the WFC3 crafted filters have considerably alleviated the contamination problem due to interlopers and provided more precise photometric redshift estimates. The standard selection method applied to these survey data sets is based on the dropout technique introduced by Steidel et al. (1996) and later constantly refined and improved by several authors (e.g. Giavalisco et al. 2004, Bouwens et al. 2007). Though this method has proved to be very solid in identifying high-redshift sources, it has the drawback that the exact source redshift cannot be determined with complete confidence. This uncertainty can be partly overcome by also using the longer wavelength infrared data, such as that provided by the {\it Spitzer} satellite; by building a more complete Spectral Energy Distribution (SED), the stellar mass, age and redshift of a given source can be constrained further. Other complementary techniques to search for distant ($z > 5$) galaxies are also widely used, among which the narrow-band spectroscopy 
(Malhotra et al. 2005; Shimasaku et al. 2006; Taniguchi et al. 2005; Kashikawa et al. 2006)
aimed at detecting the Ly$\alpha$ line, carrying a large fraction of the bolometric luminosity, is definitely the most established one. Such narrow-band searches have yielded the record-holding most distant galaxy at $z=6.96$ (Iye et al. 2006). Finally, another series of experiments involve searching for remote galaxies behind foreground galaxy clusters acting as magnification lenses (Schaerer \& Pell\'o 2005; Richard et al. 2008; Bradley et al. 2008). Although these searches result in deeper magnitudes, their interpretation is hampered by the lens modelling and by the extremely narrow field of views, rendering it difficult to keep cosmic variance under control. As a final remark, we note that the most distant, spectroscopically confirmed, cosmic object is a Gamma Ray Burst (GRB090423 at $z=8.2$, Salvaterra et al 2009b; Tanvir et al. 2009). Although not a galaxy, the presence of this indicator implies that star formation was already well under way at those early epochs, thus further encouraging deeper galaxy searches. In addition, the GRB can be seen as a signpost of the underlying galaxy which could possibly be detected in the future knowing its exact position. Such a finding would be of the utmost importance as GRBs are mostly associated with star forming dwarf galaxies (Savaglio, Glazebrook \& Le Borgne 2009) which are now considered to be the dominant sources of (re)ionizing photons at high redshifts (Choudhury \& Ferrara 2007; Choudhury, Ferrara \& Gallerani 2008).

What have we learned from this wealth of experimental results ? The most solid piece of information that can be determined from the data appears to be the luminosity function (LF) and, less robustly, its evolution. It is useful to briefly recap the present observational situation marching towards increasing redshift. Bouwens et al (2007) present a comprehensive view of galaxy candidates from the UDF/ACS/GOODS fields using NICMOS in the redshift range $z=4-6$. They identify 1416 (627) V-dropouts ($i$-dropouts) corresponding to $z\approx 5$ ($z\approx 6$) down to an absolute UV magnitude of $M_{UV}\approx -17$ with a LF described by a Schechter function with characteristic luminosity and faint-end slope given by $M_{UV}^* = -20.64 \pm 0.13$ and $\alpha=-1.66\pm 0.09$ ($M_{UV}^* = -20.24 \pm 0.19, \alpha=-1.74\pm 0.16$) respectively. The same group (Bouwens et al. 2008) has extended the data analysis to include $z\approx 7$ $z$-dropouts (8 candidates at $z=7.3$) and J-dropouts (no candidates at $z \approx 9$). More recently, the installation of WFC3 on board the HST has triggered a new series of searches. Oesch et al. (2010) used data collected during the first-epoch WFC3/IR program  (60 orbits) in the Y, J, H bands reaching a magnitude limit of AB$\approx 29 (5\sigma)$. They identify 16 $z$-dropouts in the redshift range $z=6.5-7.5$ from which they obtained a LF with ($M_{UV}^* = -19.91 \pm 0.09, \alpha=-1.77\pm 0.20$), essentially confirming the previous findings while extending it to fainter luminosities ($M_{UV} \approx -18$). Bouwens et al. (2010a) pushed the investigation to $z=8.0-8.5$ by using 5 Y-dropouts. 
Finally, Bouwens et al. (2009) were able to identify three J-dropouts. If confirmed, these sources would be the most distant objects detected so far. Similar studies using the same data has been performed by Bunker et al. (2010), who find a comparable number of $z-$ and $Y-$dropouts. McLure et al. (2010) did not apply specific color cuts as in the previous works, thus finding a larger number of candidates; however, they pointed out that about 75\% of the candidates at $z>6.3$ (100\% at $z>7.5$) allow a $z<2$ interloper solution. 
A recent analysis
of the three HUDF and of the deep ($\sim 27.5$ AB mag), wide-area ($\sim 40$
arcmin$^2$) WFC3 Early Release Science reveals
66 and 47 candidate galaxies at $z=7$ and $z=8$, respectively (Bouwens et al.
2010b). After carefully
modelling the selection volume of each field and of the possible 
contamination by spurious sources, the LF obtained from these data, while 
consistent with previous derivations of $M_{UV}^*$ and of the normalization, shows
a steeper faint-end slope with $\alpha=-1.94\pm 0.24$ and $\alpha=-2.00\pm 0.33$
at $z=7$ and $z=8$, respectively.

Besides the LF, tentative information on the physical properties of these sources can be extracted from their SED, exploiting available {\it Spitzer} data (Eyles et al. 2005; Yan et al. 2006; Stark et al. 2009). In a recent study Labb\'e et al (2010b), based on follow-up {\it Spitzer/IRAC} observations, analyzed the SED of 12 $z$-dropout and 4 Y-dropout candidates. None of them is detected in the {\it Spitzer/IRAC} 3.6 $\mu$m band to a magnitude limit of AB=26.9 $(2\sigma)$, but a stacking analysis reveals a robust detection for the $z$-dropout sample and a strong upper limit for the Y-dropout one. The stacked SEDs are consistent with a stellar mass of about $10^9 M_\odot$, no dust reddening, sub-solar metallicity, and best-fit ages of about 300 Myr, implying a formation epoch $z\approx 10$. These results for the stacked sample should be compared with those obtained by Finkelstein et al (2010) who performed an object-by-object analysis and found similar ages but with a considerable spread, allowing ages as low as a few Myr.       

One of the major triggers to look for very high-$z$ galaxies is the quest for the reionization sources. The ionizing photon budget provided by the candidate high-$z$ galaxies is often estimated by extrapolating their LF to lower luminosities, a step that introduces a considerable uncertainty in the final determination. Having this in mind, it is still interesting to note that most studies tend to agree on the fact that the integrated UV specific luminosity for the detected galaxies at $z=7-8$ falls short of accounting for the ionizing power required to reionize the intergalactic medium. Of course, this conclusion is subject to at least two major unknown factors, these being the gas clumping factor (affecting its ability to recombine), and the escape fraction of ionizing photons (affected by dust and neutral hydrogen absorption within galaxies). Additionally, the effect of poorly constrained ages and metallicities (including the presence of metal-free, massive Pop III stars), further complicate the calculation.

In spite of the large experimental effort, surprisingly little attention has been devoted by modelers to the very high redshift universe. Most of the work has so far concentrated on a 
semi-analytical approach (Stiavelli, Fall \& Panagia 2004; Schneider et al. 2006; Bolton \& Haehnelt 2007; Mao et al. 2007; Samui, Subramanian, Srianand 2009; Trenti et al. 2010) to compute the luminosity function, number counts and emissivity evolution of high-$z$ galaxies. Albeit quite fast and versatile, these methods cannot provide detailed information on the properties of the galaxies, often being based on simplified assumptions. Numerical dedicated simulations attempting to model galaxy populations beyond $z=5-6$ are also very scarce, with the partial exceptions constituted by 
the works by Nagamine et al. (2006) and Finlator, Dav\'e \& Oppenheimer (2007). 

Our approach is novel and different in spirit from all the previous theoretical ones. As our main aim is to model very high redshift reionization sources, we can afford smaller simulation boxes, thereby reaching the high resolutions required to resolve the dominant reionization sources - dwarf galaxies. Most importantly, though, we have implemented a careful treatment of metal enrichment and of the transition from Pop III to Pop II stars, along with a careful modelling of supernova feedback. Here we are interested in deriving the LF plus other observables from the simulations and to cast them in a form that can be compared directly with the available data or used to make new predictions for the James Webb Space Telescope (JWST).

\section{Numerical simulations}
\label{simulations}
For the present study we have performed a set of cosmological\footnote{Throughout the paper, we adopt a $\Lambda$CDM cosmological model with parameters $\Omega_M = 0.26$,
  $\Omega_{\Lambda} = 0.74$, $h=0.73$, $\Omega_b=0.041$, $n=1$ and
  $\sigma_8=0.8$, in agreement with the 3-yr WMAP results (Spergel et
  al. 2007).} simulations using the publicly available code GADGET\footnote{www.mpa-garching.mpg.de/galform/gadget/} (Springel 2005) with an improved treatment of chemical enrichment as detailed in Tornatore, Ferrara \& Schneider (2007, TFS07). A unique feature of the computation concerns the Initial Mass Function (IMF) of stars, which is taken to
be different for Pop III and Pop II stars and depends on gas metallicity, $Z$.  In brief, if $Z < Z_{\rm cr}$, the adopted IMF is a Salpeter 
law with lower (upper) limit of $100M_\odot$ ($500 M_\odot$) and only pair-instability supernovae  ($140 M_\odot < M < 260 M_\odot$) contribute to metal enrichment
(Heger \& Woosley 2002). If $Z \ge Z_{\rm cr}$, the
above limits are shifted to $0.1 M_\odot$ ($100 M_\odot$),
respectively; stars above $40 M_\odot$ end their lives as black holes
swallowing their metals. According to the canonical choice, we fix $Z_{\rm
  cr} = 10^{-4}Z_\odot$. These two populations, to which we will refer
to as Pop III and Pop II stars respectively, differ also for their
metal yield and explosion energy. Complete details of the simulation can be found in TFS07. The simulation follows
the production and transport of six different metal species, namely:
C, O, Mg, Si, S, Fe, based on which the locally appropriate IMF is selected.  
The simulated volume has a linear (comoving) size $L=10 h^{-1}$~Mpc with 
$N_p=2\times 256^3$ (dark+baryonic) particles, corresponding to a dark matter 
(baryonic) particle mass of $M_p=3.62 \times 10^6 h^{-1} M_\odot$ ($6.83 
\times 10^5 h^{-1} M_\odot$); the corresponding force resolution is $2~$kpc. Our
resolution does not allow us to track the formation of mini-halos, whose stellar 
contribution remains very uncertain due to radiative feedback effects (Haiman 
\& Bryan 2006; Susa \& Umemura 2006; Ahn \& Shapiro 2007, Salvadori \& Ferrara 2009).
The computation is initialized at $z=99$ and carried on until $z=2.5$. 
Star formation is treated as a stochastic process, following Springel
\& Hernquist (2003) Basically, at a given time the star
formation rate of a (multi--phase) gas particle is computed using a
Schmidt-type law $\dot m = xm/t_*$, where $xm$ is the mass of cold
clouds providing the reservoir for star formation. Within the
effective star formation model by Springel \& Hernquist (2003), the star formation time-scale,
is computed as $t_*(\rho) = t_*^0 (n/n_*)^{-1/2}$, where $t_*^0=1.5$
Gyr and the density threshold for star formation, $n_* = 0.1$
cm$^{-3}$, are free parameters of the model chosen so as to reproduce
the observed Kennicut relation within uncertainties.

Supernova winds are treated as in the original model by Springel \&
Hernquist (2003); however,
for simplicity and because the mass load and kinetic energy fraction
are unknown parameters, we have given the winds from both populations
an initial velocity $v_w = 500$ km s$^{-1}$, which appears to be
consistent with that derived from observations of high-z starburst
galaxies. Wind particles are temporarily hydrodynamically decoupled
until either (i) they have moved by a traveling length $\lambda = 2$
kpc, or (ii) their density has decreased below 0.5 times the star
formation density threshold. The metals are donated by star particles
to the surrounding gas ones using a SPH kernel as described in
TFS07.

The gas photo-ionization and heating rates are calculated at equilibrium with a background ionizing radiation due to the combined contribution of galaxies and quasars, taken from Haardt 
\& Madau (1996), shifted so that the intensity at 1 Ryd is $J_\nu= 0.3\times
10^{-21}$~erg ~s$^{-1}$~Hz$^{-1}$, in agreement with Bolton et al.
(2005). Gas cooling and wind treatment details are the same as in TFS07. 

\begin{figure*}
\center{\includegraphics[scale=0.9]{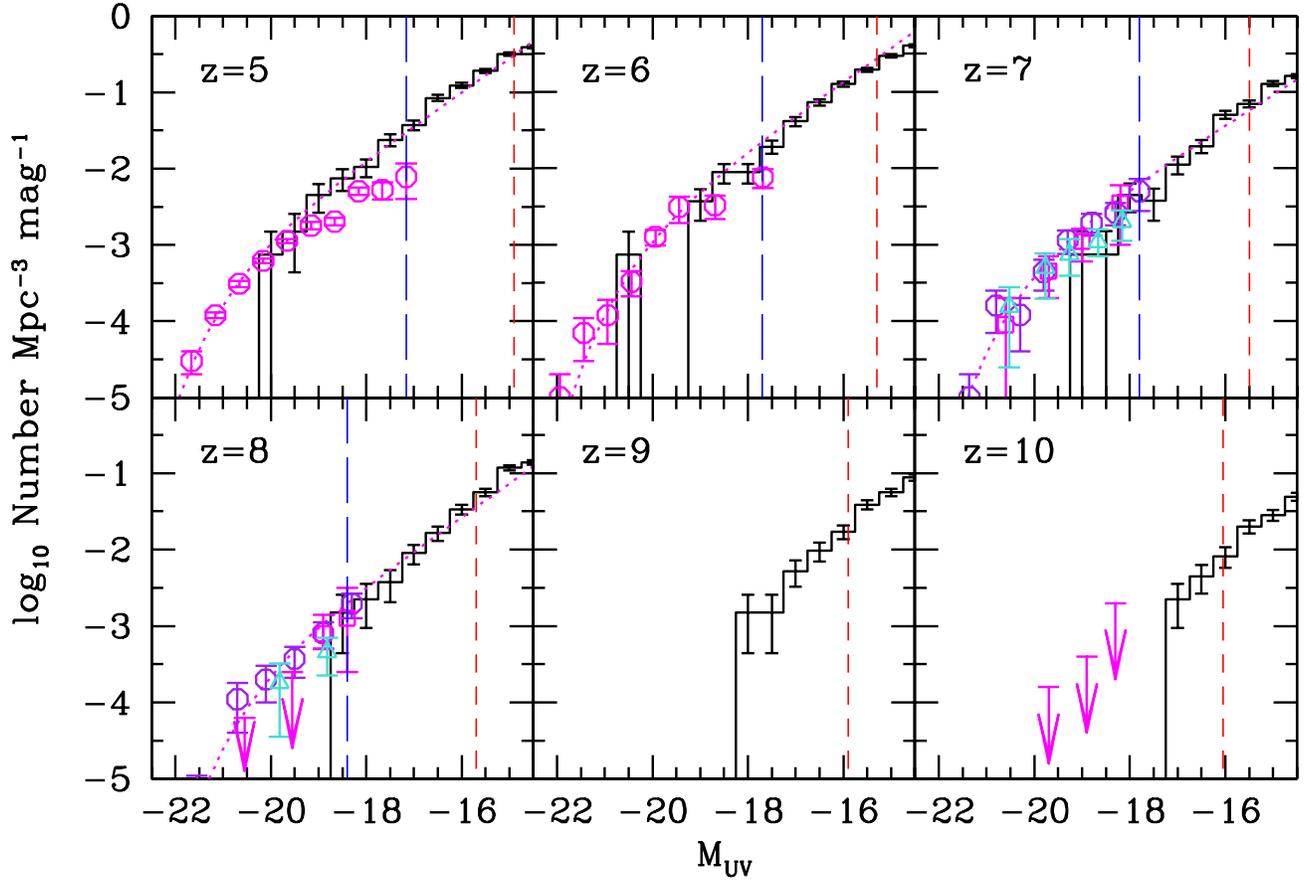}}
\caption{\label{fig:LF} The UV LF of galaxies at the different redshifts shown in 
each panel. Observational data (and upper limits) from HUDF are taken from Bouwens et al. (2007) for $z=5,6$; Oesch et al. (2010; squares), McLure et al. 
(2010; triangles) and Bouwens et al. (2010b; circles) for $z=7$; 
Bouwens et al. (2010a; squares), McLure et al. (2010; triangles) and Bouwens
et al (2010b; circles) for $z=8$ and Bouwens et al. (2009) for $z=10$; they are shown as circles (arrows). The histograms show the 
simulated LF with error bars representing Poisson errors. Dotted lines are the
Schechter function fits to the LF;  the vertical short(long)-dashed lines mark the sensitivity limit of JWST (HST/WFC3).
}
\end{figure*}
\section{Luminosity Function}

As a first check of the simulation results we will compute the evolution of the 
LF of the simulated galaxies and compare it with available data. To this end we make a number of physical assumptions that are discussed in the following. 

For each galaxy at a redshift $z$, the luminosity at rest-frame wavelength, $\lambda$, is the sum of the contribution of its Pop II and Pop III stars. The SED of Pop II stars is computed by running the Starburst99 code (Leitherer et al. 1999, Vazquez \& Leitherer 2005) using 
the metallicities and stellar ages appropriate for the galaxy under consideration as obtained from the cosmological simulation. The SEDs of massive (metal-free) Pop III stars have been computed by Schaerer (2002) including the effect of nebular emission. For both populations the IMF is taken according to the same prescription used in the simulation and described in Sec. \ref{simulations}. The total luminosity of a galaxy is then given by

\begin{equation}
L_\lambda=L_{\lambda}^{\rm II}+L_{\lambda}^{\rm III}=
l_{\lambda}^{\rm II}(\tau^{\rm II},Z)\dot{M_*}^{\rm II}+l_{\lambda}^{\rm III}\dot{M_*}^{\rm III}\tau^{\rm III}
\end{equation}

\noindent
where $l_{\lambda}^{\rm II}(\tau^{\rm II},Z)$ is the SED template for Pop II stars with mean age $\tau^{\rm II}$ and metallicity $Z$ corresponding to a continuous star formation rate of 1 $\Msun$ yr$^{-1}$; $\dot{M_*}^{\rm II}$ is the Pop II star formation rate. The SED template of Pop III stars, whose star formation rate is $\dot{M_*}^{\rm III}$, is $l_{\lambda}^{\rm III}$; 
$\tau^{\rm III} = 2.5\times 10^6$ yr is the mean lifetime of massive Pop III stars (Schaerer 2002). We have implicitly assumed that the emission properties of Pop III stars are constant during the short lifetime of these massive metal-free stars. Next, the luminosity is converted into absolute AB magnitudes, $M_{UV}$, where the suffix UV refers to the wavelength $\lambda= (1600, 1350, 1500, 1700, 1500, 1600)$ \AA~for $z=(5, 6, 7, 8, 9, 10)$, respectively. Though this wavelength choice is chosen to be consistent with the observational data we compare with, a different value would not affect the results in any sensible way, given the flatness of the spectrum in this short wavelength range.

\begin{figure*}
\begin{tabular}{lr}
\includegraphics[scale=0.32]{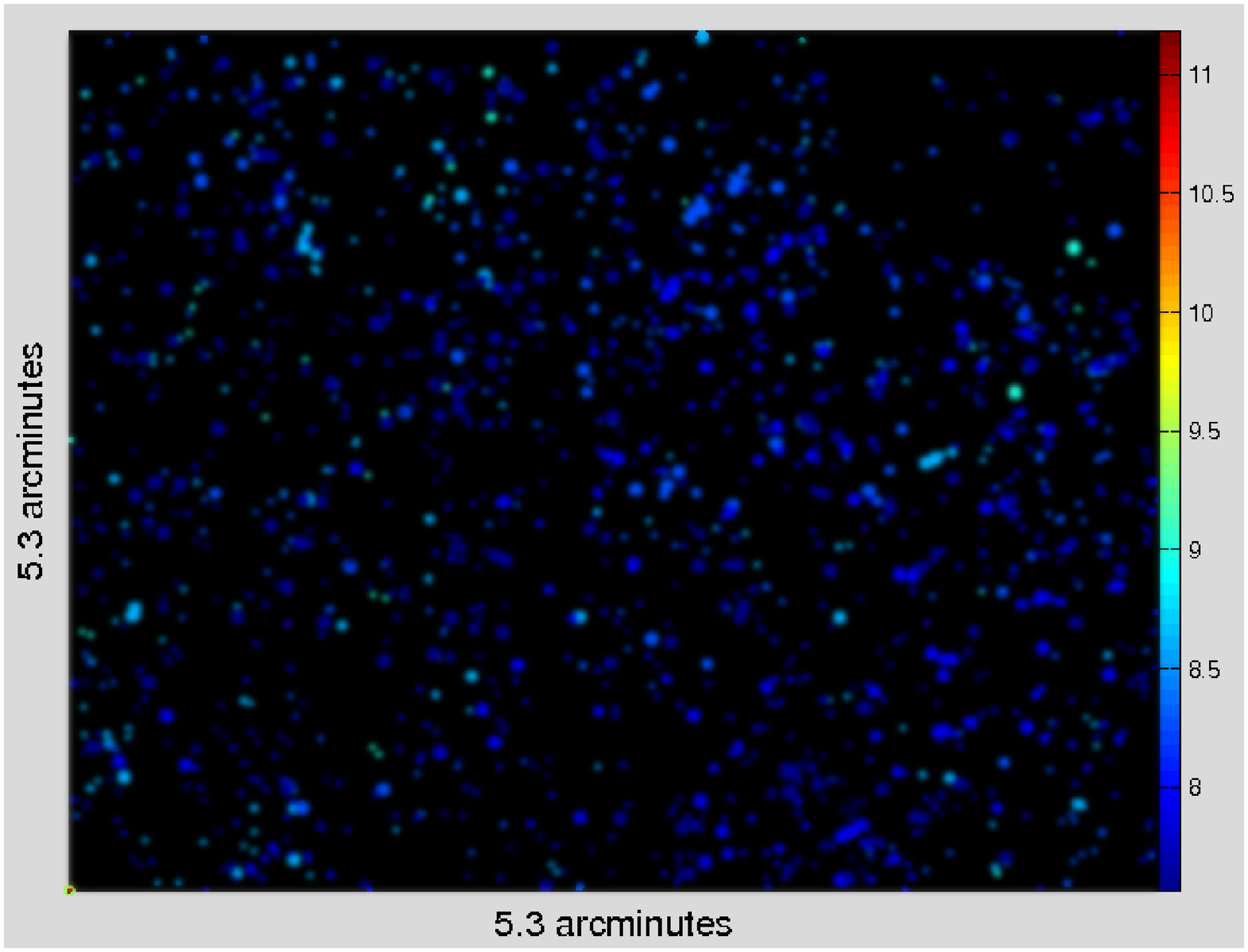}& \includegraphics[scale=0.32]{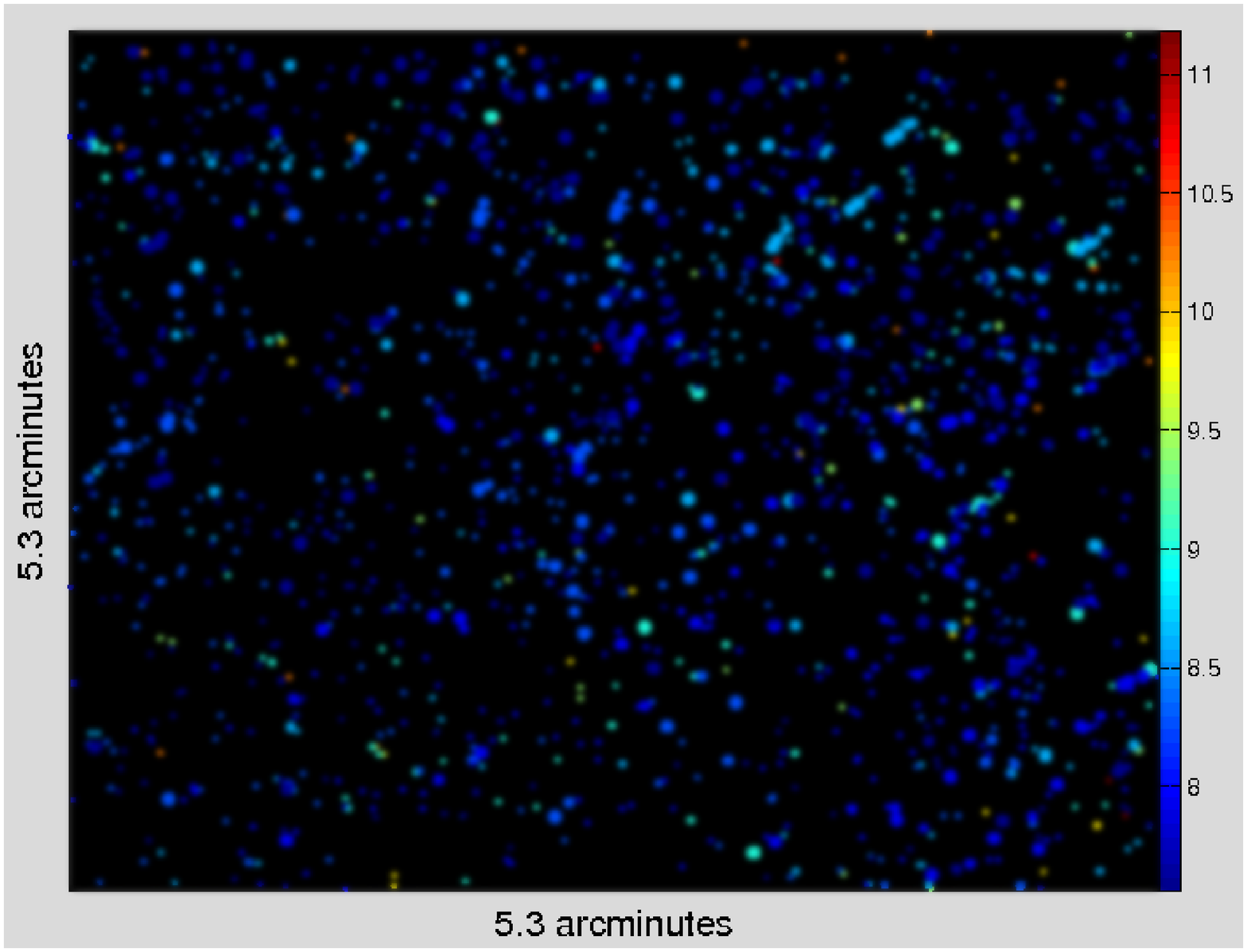}
\end{tabular}
\caption{Maps showing the distribution of galaxies with an observed flux larger than the JWST sensitivity limit of 1 nJy in the J-band (left panel) and H-band (right). The maps are a 2D cut of a 3D image produced by stacking simulation snapshots between $z \sim 7.6-11.6$. The vertical color bar gives the redshift of the object, the size of the galaxies scales with their flux in the range 1 nJy $< {\cal F} <$ 24.4 nJy.}
\label{maps}
\end{figure*}

The LF of galaxies at any redshift $z$ is obtained by counting the galaxies with a given absolute magnitude in each magnitude bin and dividing the final result by the total volume of the
simulation and bin size (0.5 mag). We perform this procedure for the following six redshifts $z=(5, 6, 7, 8, 9, 10)$.  The results are shown in Fig.~\ref{fig:LF} as solid histograms, where the error bars represent the Poisson error on the number of galaxies in each magnitude bin. These theoretical LFs are then compared to the experimental ones collected from the various analyses of the HUDF. For $z=10$ we show the upper limits on the LF obtained from the three available candidates identified by Bouwens et al. (2009). 

Let us now analyze the results shown in Fig. \ref{fig:LF} in more detail. It is clear that the luminosity range sampled by the observations and our predictions is only partially overlapping. This is because on one hand, even the exquisite sensitivity of WFC3 is not sufficient to properly sample the faint-end of the LF ($M_{UV}\simgt -18$); on the other hand, our simulations, which are specifically designed to properly resolve the very first galactic units in a relatively small volume, lack the most massive, rare objects which comprise the bright end of the LF. In spite of these shortcomings, we consider it a rewarding success that the amplitudes of the theoretical and experimental LFs match almost perfectly, and at the same time, have quite similar slopes at all redshifts for which data are available. This is even more striking as no attempts have been made to fit or adjust the theoretical curves to the observed LF, i.e. they have been computed directly from the simulation output with no free extra parameters.

\begin{figure}
\center{\includegraphics[scale=0.43]{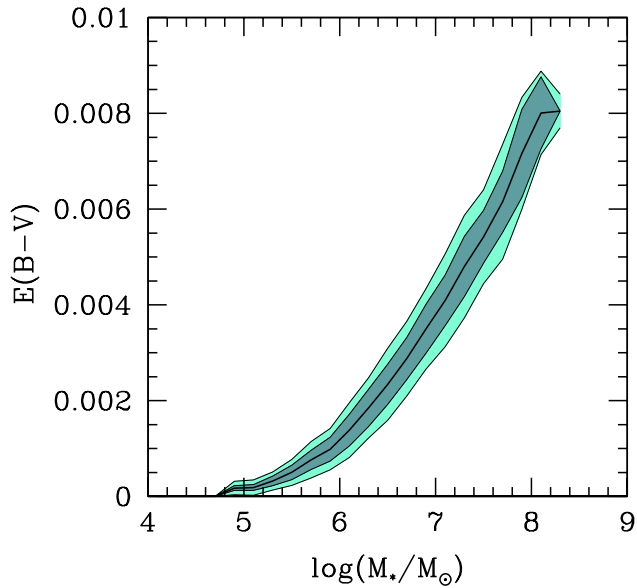}}
\caption{the color excess, $E(B-V)$, as a function of the stellar mass for $z\approx 7$ galaxies obtained from the simulation snapshot. The dark (light) region shows the color excess for 65\% (95\%) of the galaxies. }
\label{ebmv}
\end{figure} 

Our results suggest two clear trends. First, the LFs shift towards fainter luminosities with increasing redshift, mimicking a pure luminosity or density evolution. This is quite consistent with the trend of an increasing $M^*_{UV}$ with redshift, found in the data by several groups (see the extended discussion in Ouchi et al. 2009), preferring a pure luminosity evolution. Second, the faint-end slope of the simulated LF does not vary (within errors) from $z=5$ to $z=10$, maintaining an almost constant value of $\alpha=-2.0$. This value is slightly larger than the one derived from the data (see Sec. \ref{introduction}) at
$z=5-6$ but perfectly consistent with the recent determination of $\alpha$ for
the $z=7-8$ LF obtained by Bouwens et al. (2010b). This behavior is most likely produced by a combination of the halo mass function evolution, feedback effects and evolving stellar populations.   

Clearly the faint-end slope will be better constrained by forthcoming facilities such as the JWST; the above successful test of our model allows us to make reliable predictions for surveys that will be performed with such instruments. For a deep exposure of $10^6$ s, JWST is expected to reach a photometric sensitivity of about $1$~nJy ($10\sigma$) allowing an investigation of the faint-end of the LF predicted by our simulations at least up to $z=9$ (see Fig. \ref{fig:LF}). This will be particularly exciting because it will very likely unveil the physical properties of these objects which are now thought to be the main reionization sources, as we will discuss in detail in Sec. \ref{Rei_sources}. 
We want to stress here that our simulation allow us to resolve properly (i.e. with more
than 200 DM particles) the galaxy population
that will be observed by JWST, being their total halo masses $\ge 10^9\;\Msun$.
This, toghether with the carefull threatment of the chemical feedback, makes
our predictions particularly robust.

To give a visual idea of the high-$z$ ($z > 7.5$) galaxy population in a typical JWST field (of size 5.3 arcmin) we have produced the maps shown in Fig. \ref{maps} which demonstrate the wealth of objects that JWST will be able to detect up to $z \simgt 11$. The maps are a 2D cut of a 3D image produced by stacking all the available simulation snapshots between $z \sim 7.6-11.6$, and filling in the gaps between snapshots by randomly rotating the orientation axes in all the three-dimensions simultaneously. 

As it has emerged from the overview of the observational results given in the Introduction, dust seems to have only a minor effect on shaping the SED of faint high-redshift galaxies. This is easily understood, as we will see later on, as a result of the relatively low stellar masses and metallicities characterizing these objects. Dust can however become more important in luminous Lyman Break Galaxies (LBGs) and Lyman Alpha Emitters (LAEs) as many studies have shown (Lai et al. 2007; Atek et al. 2008; Nagamine, Zhang \& Hernquist 2008; Dayal et al. 2008, 2009, 2010, 2011; Finkelstein et al. 2009). 

The dust enrichment of galaxies can be followed self-consistently by post-processing our simulation through the model introduced by Dayal et al. (2010), using which, we can quantitatively check the previous statement. In brief, the model assumes that dust is produced by supernovae (we neglect here the contribution of AGB stars, though see Valiante et al. (2009) who found a significant contribution from these objects for population ages  $>150-200$ Myr)  
%can be neglected for $z \ge 5.7$, since the typical evolutionary timescale of these stars ($> 1$ Gyr) becomes longer than the Hubble time) 
and taking into account three processes: (a) dust forms in the expanding ejecta with a yield per SNII of  $0.54 M_\odot$ (Todini \& Ferrara 2001; Nozawa et al. 2003; Bianchi \& Schneider 2007), (b) SNII destroy dust in the ISM they shock to velocities $> 100 $ km s$^{-1}$, with an efficiency of 0.12 (McKee 1989), and (c) a homogeneous mixture of gas and dust is assimilated
into star formation (astration). Once the dust mass is calculated for each galaxy in the simulation, we translate this into a value of $E(B-V)$ using the appropriate SN dust extinction curve given by Bianchi \& Schneider (2007) as explained in Dayal et al. (2010). 
The resulting values of $E(B-V)$ for galaxies at $z\approx 7$ are shown in Fig. \ref{ebmv}
as a function of the stellar mass. Many galaxies, especially the smallest ones, are
almost dust-free, and none of them shows a dust reddening value larger than  $E(B-V)=0.009$. This
evidence allows us to safely neglect the effects of dust on the UV LF.

\section{Properties of first galaxies}

\begin{figure*}
\center{\includegraphics[scale=0.78]{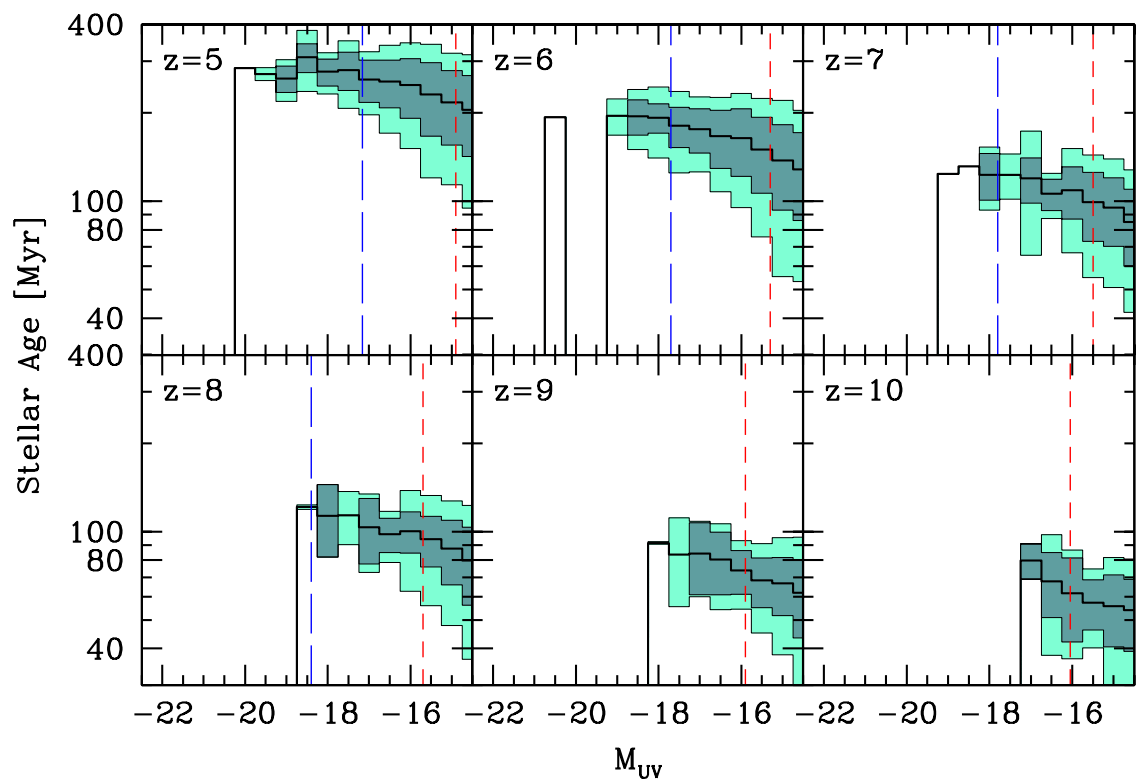}}
\caption{\label{fig:age} Mean stellar age of galaxies located at different redshifts 
(see labels) as a function of their absolute UV magnitude. The dark (light) shaded 
area show 68\% (95\%) of the galaxies in the magnitude bin.  The vertical short (long)-dashed lines mark the sensitivity limit of JWST (HST/WFC3).
}
\end{figure*}

\begin{figure*}
\center{\includegraphics[scale=0.78]{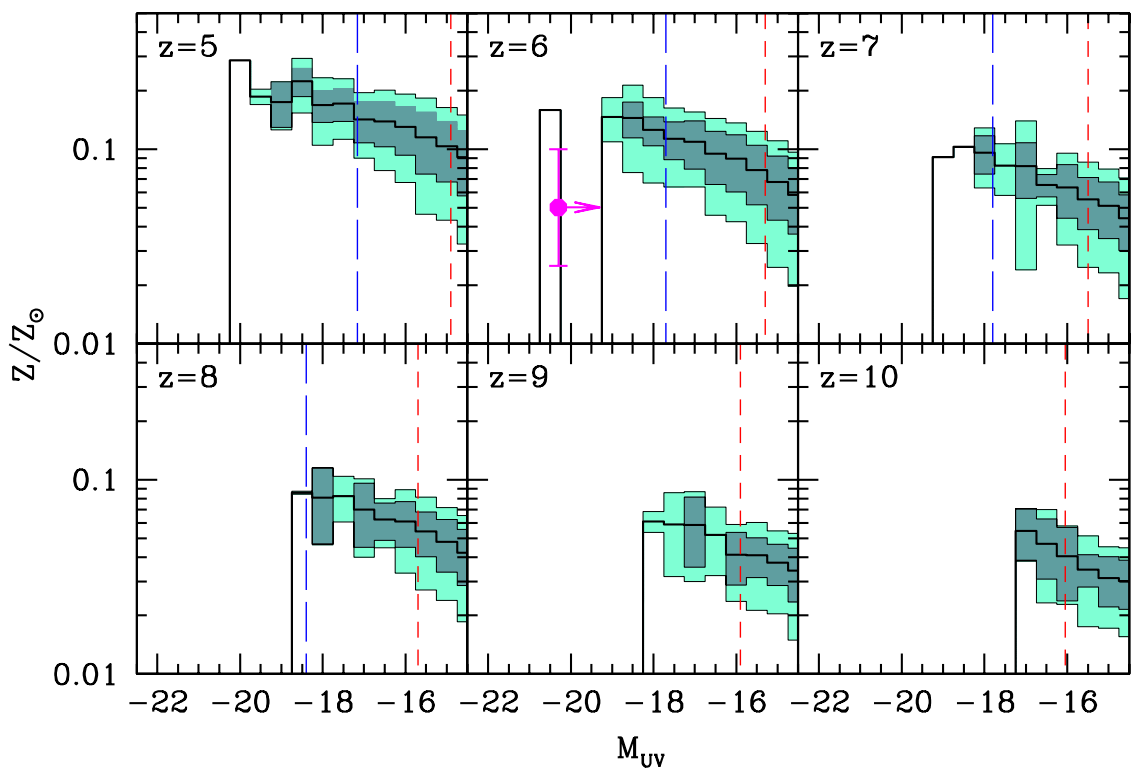}}
\caption{\label{fig:metal} Mean metallicity of galaxies located at different redshifts 
(see labels) as a function of their absolute UV magnitude. The dark (light) shaded 
area show 68\% (95\%) of the galaxies in the magnitude bin. The vertical short (long)-dashed lines mark the sensitivity limit of JWST (HST/WFC3).
 The filled circle refer to the GRB~050904 host galaxy at $z=6.3$.}
\end{figure*}
In addition to predicting the global evolution of the LF, a major strength of our study is that it makes possible to extract the physical properties of the high-$z$ galaxies which are a part of the  faint-end of the LF. We start by concentrating on the stellar ages, reported in Fig. \ref{fig:age} as a function of the UV magnitude, from which the following main conclusions can be drawn. On average, fainter (and less massive) objects tend to be younger at all redshifts, with typical ages for observable objects in the range 200-300 Myr at $z=5$ and 80-130 Myr at $z=7-8$, with the caveat that a considerable age spread exists at all luminosities. These ages imply that these galaxies started to form stars as early as $z=9.4$, clearly suggesting that their UV light might have influenced the cosmic reionization history. From a different perspective, we note that stellar ages of $\sim 100-150$ Myr are expected for galaxies at the sensitivity limit of WFC3 for $z=7-8$, in agreement with the observational estimates obtained by Labb\'e et al. (2010a) and Finkelstein et al. (2010).

A second important key physical parameter of pristine galaxies is their stellar and gas metallicity. In the following we will only discuss the stellar metallicity, keeping in 
mind that the two closely match each other. Somewhat surprisingly (but not unexpectedly) 
even the faintest galaxies appear to be already enriched to remarkable levels: at the JWST sensitivity threshold, we find $Z > 0.03 Z_\odot$ at all redshifts; galaxies identified 
by HST ($M_{UV} < -18$) systematically show metallicities in excess of $Z=1/10 Z_\odot$ even at $z=7-8$. Thus we come to the interesting conclusion that even at these early epochs, the self-enrichment, due to the metals produced, following the first star formation episodes is able to increase the metal abundances of such small objects to levels comparable to their present-day counterparts (e.g. the Magellanic Clouds). In addition, such high mean metallicites could in principle preclude the formation of Pop III stars according to the critical metallicity scenario (Schneider et al 2002, 2003; Schneider \& Omukai 2010) which predicts $Z_{cr}=10^{-5\pm 1} Z_\odot$ as the upper limit 
for the formation of Pop III stars. However, as we will discuss in more detail later, inside these early structures, small pocket of (quasi) pristine gas may survive in which a relatively tiny amount of Pop III stars continue to form as pointed out by TFS07 and Jimenez \& Haiman (2006). In brief, the scenario leads to the concept of a "PopIII wave", i.e. the physical
phenomenon by which in each galaxy the formation of stars below the
critical metallicity is progressively segregated towards the external
regions of the galaxy, where almost unpolluted regions are still
present. Until this process comes to an end (when metal pockets
produced around the first stars forming regions reach a considerable
volume filling factor in the system), PopIII and PopII formation modes
coexist in the same galaxy. One might naively think that this process
can take longer in large galaxies than in small ones, but it is
(roughly) true only to zero-th order approximation, as in lower mass
galaxies the efficiency of star formation is also depressed and the
amount of time required, once scaled by the baryonic mass of the
system, is not dramatically different from that of larger systems.

Taken together, Fig.~\ref{fig:age} and \ref{fig:metal} provide a first guess 
of the properties of a typical high-$z$ galaxy with given
absolute UV magnitude $M_{UV}$. This can be useful when comparing the observed
photometric data  with synthetic SEDs.

\begin{figure}
\center{\includegraphics[scale=0.43]{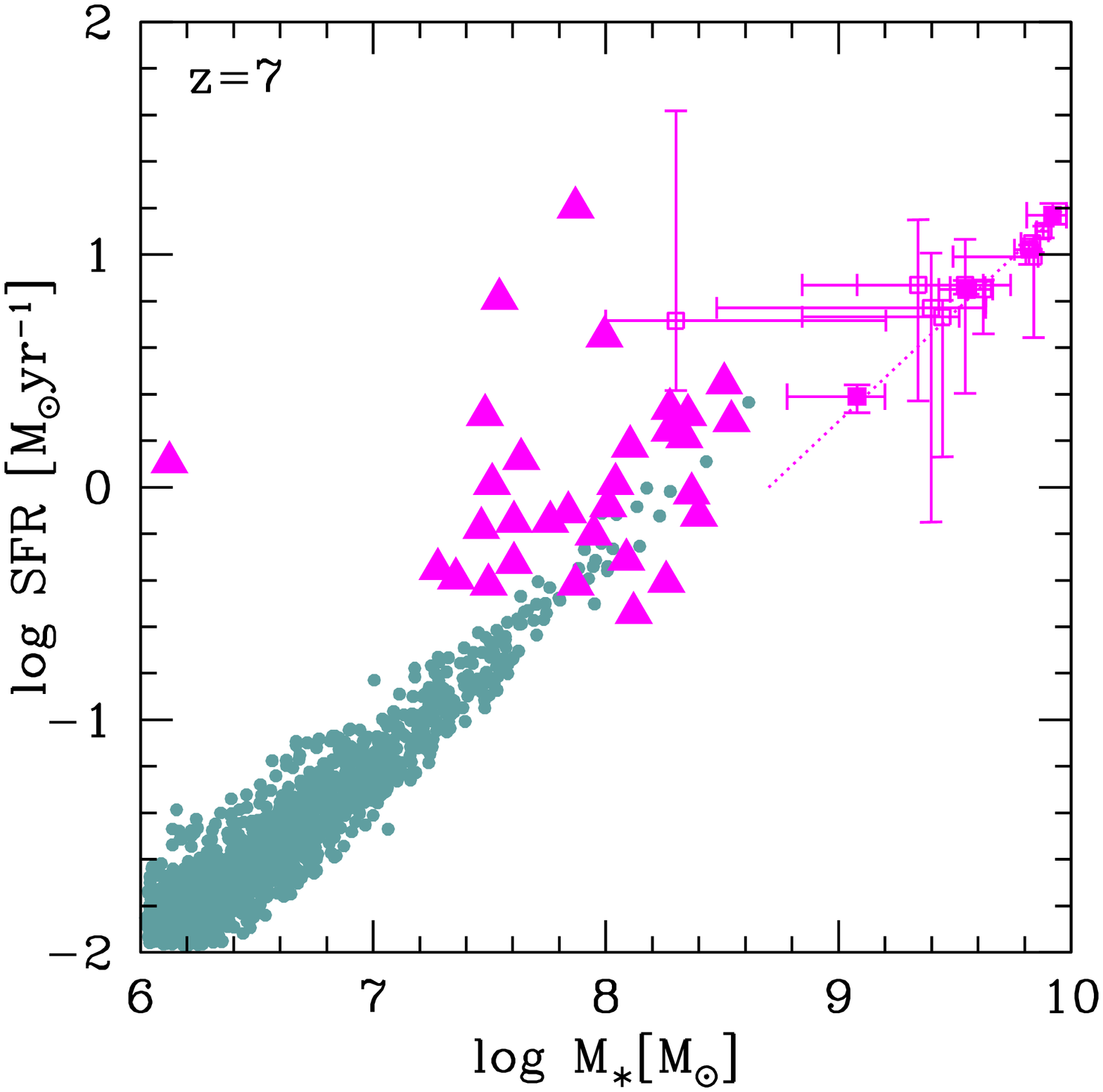}}
\caption{\label{fig:schaerer} Stellar mass--SFR relation for simulated 
galaxies at $z=7$ (black points) compared with recent estimates for observed
$z$-dropout as derived 
by different authors: triangles are from Schaerer \& de Barros (2010), 
open squares from Gonzales et al. (2009), and filled squares for the mean
values derived by Labb{\'e} et al. (2010b). The dotted line shows the empirical
relation between the two quantities obtained by Labb{\'e} et al. (2010a).}
\end{figure}

Fig.~\ref{fig:schaerer} shows the relation between the SFR and stellar mass of
galaxies at $z=7$. Simulated objects follow an almost linear relation with 
significant spread, i.e. an almost constant specific star formation rate. 
This trend closely matches the one found in the
analysis of stacked SEDs of WFC3 $z$-dropout by Labb{\'e} et al. (2010b), 
although a single object analysis reveals large errors in the determination
of both the stellar mass and the SFR for these objects (Gonzalez et al. 2010). 
We note that simulated galaxies tend to have a slightly higher SFR (for a fixed 
stellar mass) than expected by the extrapolation of the relation to
smaller objects. However, in a re-analysis of the 
observed $z=7$ sample including the effect of nebular continuum and line 
emission along with that of dust absorption, Schaerer \& de Barros (2010) find smaller stellar masses
and larger SFR with respect to previous works. Our simulated galaxies lie 
somewhat in between these two observational estimates.

\begin{figure*}
\center{\includegraphics[scale=0.9]{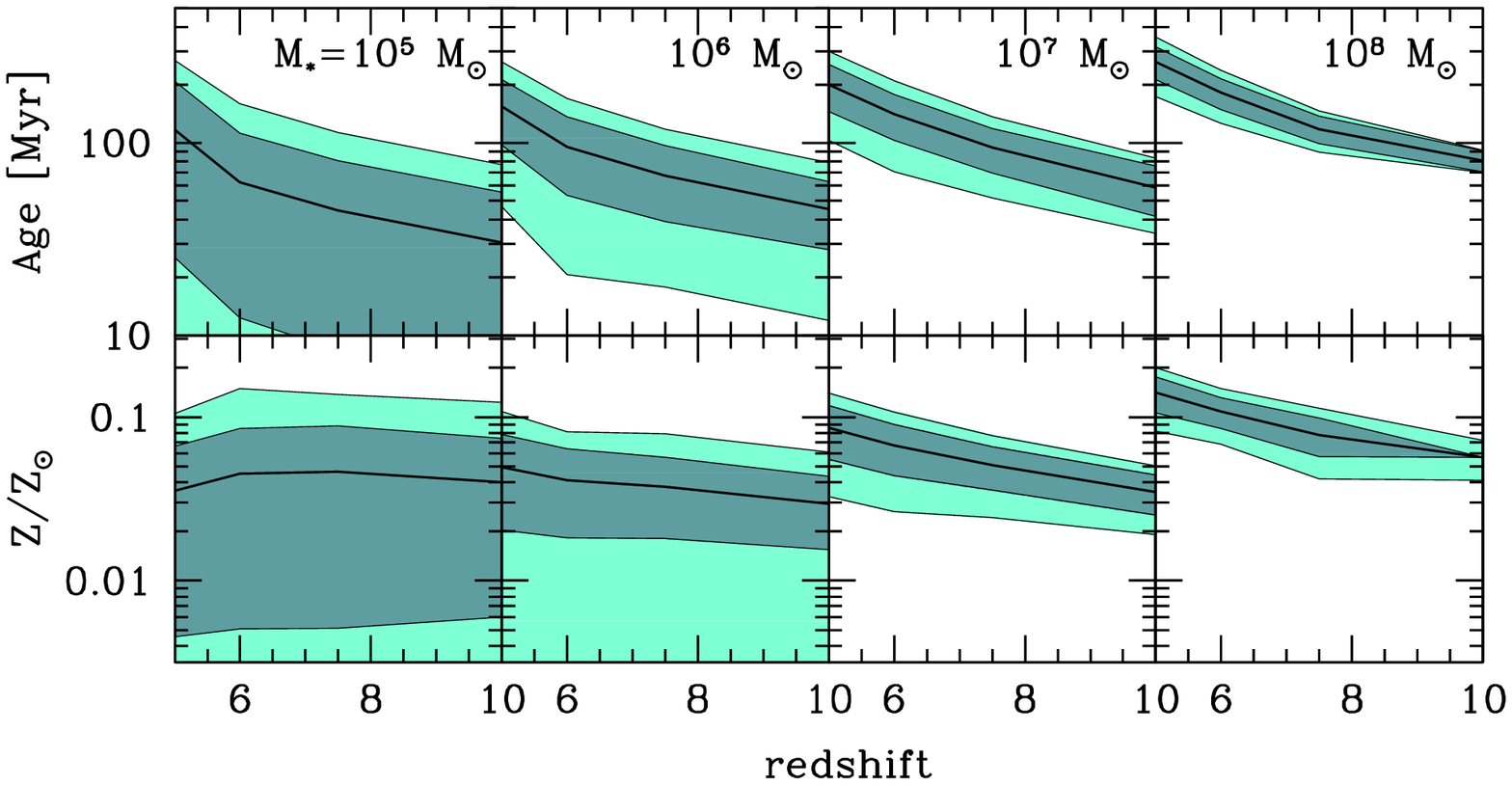}}
\caption{\label{fig:prop_mass} Redshift evolution of the mean stellar age and 
metallicity for galaxies in given stellar mass ranges (see labels). 
The dark (light) shaded area show 68\% (95\%) of the galaxies in the redshift bin. 
}
\end{figure*}

Fig.~\ref{fig:prop_mass} shows the evolution of the galaxy properties as a
function of redshift for galaxies with different stellar masses. 
The mean age of stellar population for all galaxies in the simulation is found
to decrease with redshift $\propto (1+z)^{-2}$; however, the age spread increases 
for less massive galaxies at any redshift. The average stellar metallicity is also 
growing with time for $M_*>10^7 M_\odot$, but show a much flatter (almost constant) 
evolution in the smallest objects, leveling at about 1/25 $Z_\odot$ for the tiniest star-forming galaxies. However, even for dwarf galaxies with $M_* \approx 10^5 M_\odot$ one 
can find individual objects enriched up to $1/10$ of solar metallicity already at $z=10$.    
This mass-dependent metallicity evolution is probably caused by the different ability  
to retain metals deposited by supernova explosions of the massive and dwarf populations. 
As the potential wells of the latter one are shallower, metals escape easily into the intergalactic medium thus setting an upper limit to the amount of metals than can be 
kept in their main body (Mac Low \& Ferrara 1999). An obvious implication is that IGM 
metals preferentially come from small and common objects, thus resulting in a more 
homogeneous enrichment (Ferrara, Pettini \& Shchekinov 2000; Madau, Ferrara \& Rees 2001).

%\begin{figure}
%\center{{
%\includegraphics[scale=0.5]{metgas0.15.ps,height=8cm} }}
%\caption{\label{fig:metgas} ISM metallicity. The filled circle show the result obtained using
%observations of GRB~050904 at $z=6.3$, where the metallicity is given by 
%spectrum of the GRB afterglow (Kawai et al.) and the limit on the absolute 
%magnitude of the GRB~050904 host galaxy (Berger et al. 2007)}
%\end{figure}

\section{Reionization sources}\label{Rei_sources}
Having obtained the luminosity function and characterized some of the physical properties of the current high-$z$ candidate galaxies, we now turn to an analysis of their role in the reionization process. 

The rate of ionizing photons from the $j$-th galaxy is given by

\begin{equation}
\dot{n}_{ion,j}=f_{esc}\left[Q^{\rm II}\dot{M}^{\rm II}_{j}+Q^{\rm III}\dot{M}^{\rm III}_{j}\tau^{\rm III}\right]
\end{equation}

\noindent
where $Q^{\rm II}$ ($Q^{\rm III}$) is the ionizing photon flux for Pop II (Pop III) stars, which of course is a function of the IMF, metallicity and stellar age of any given galaxy. Finally $f_{esc}$ is the escape fraction of ionizing photons. For simplicity, we assume a redshift-independent value of $f_{esc}$ for the two stellar populations. The ionization rate provided by galaxies in the simulation box per unit comoving volume at redshift $z$, $\dot{N}_{ion}(z)$, is then given by the sum over all galaxies at that redshift divided by the volume of the simulation. 

The actual ionization rate must also include galaxies that are too rare to be caught in our relatively small simulation volume. To account for this correction, we add the ionization rate due to bright galaxies by integrating the observed LF (or the upper limits) from the luminosity of the brightest galaxy in the output down to very low magnitudes ($M_{UV}=-25$).   
The ionizing photon flux is obtained using a SED derived assuming that rare galaxies have the same stellar age and metallicity as the brightest simulated one. The extra contribution of the unaccounted bright-end of the LF is found to be at most $10$\% of the total ionizing photon emission. 

\begin{figure}
\center{\includegraphics[scale=0.42]{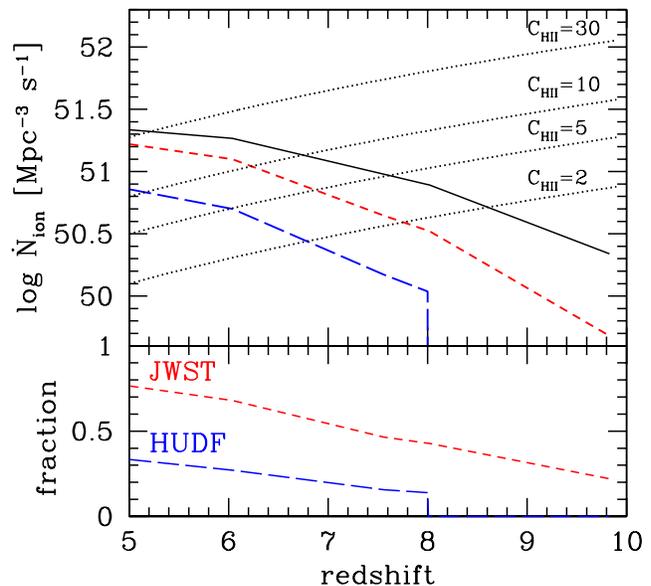}}
\caption{\label{fig:reion} {\it Upper panel:} Redshift evolution of the total specific ionization rate (solid line). The short (long)-dashed line corresponds to galaxies detectable by JWST(HST/WFC3); dotted lines show the specific IGM recombination rate for different values of the clumping factor $C_{\rm HII}$. {\it Bottom panel:} Fraction of ionizing
photons coming from galaxies identified by JWST and in the HUDF. An escape fraction $f_{esc}=0.2$ has been assumed.}
\end{figure}

The evolution of the total specific ionization rate, $\dot{N}_{ion}$, is plotted in the top panel of Fig.~\ref{fig:reion} ($f_{esc}=0.2$) along with the same rate due to galaxies detected in the HUDF, $\dot{N}_{ion}^H$, or detectable by JWST, $\dot{N}_{ion}^J$. The ratios
between ($\dot{N}_{ion}^H$, $\dot{N}_{ion}^J$) and $\dot{N}_{ion}$ are also shown for clarity in the bottom panel of the same Figure. HST is now resolving the sources that provide $\approx 1/3$ of the ionizing photon budget at $z=5$ and $\sim 20$\% at $z=7-7.5$. This results is consistent with the limits set by the
observed very steep faint-end slope of the $z=7-8$ LF (Bouwens et al. 2010b) 
and with the estimate of recent semi-analytical models 
(Choudhury et al. 2008, Trenti et al. 2010).

At the sensitivity limit of JWST, it will be possible to detect the bulk of ionizing sources up to $z\sim 7.3$, but at higher redshifts most of the ionizing photons will still be produced by sources that are too faint to be detected even by JWST. 

The total ionization rate density $\dot{N}_{ion}(z)$ should then be compared 
with the recombination rate density of the IGM, $\dot{N}_{rec}(z)$, given by (e.g. Madau, Haardt \& Rees 1999)

\begin{equation}
\dot{N}_{rec}=\frac{\langle n_H \rangle}{\langle t_{rec} \rangle}=10^{50.0}C_{HII}\left(\frac{1+z}{7}\right)^3 {\rm s}^{-1} {\rm Mpc}^{-3},
\end{equation}

\noindent
where $\langle n_H \rangle$ is the mean comoving hydrogen density in the Universe and 
$\langle t_{rec} \rangle$ is the volume-averaged recombination time for ionized hydrogen
with an effective H$_{\rm II}$ clumping factor $C_{HII}=\langle n^2_{HII}\rangle/\langle n_{HII}\rangle^2$. The recombination rate density is shown in the top panel of Fig.~\ref{fig:reion} with dotted lines for different value of the clumping factor $C_{HII}$. For $f_{esc}=0.2$ and $C_{HII}=10$, the balance between ionization and recombination is obtained 
at $z\sim 6.8$. For $f_{esc}=0.1$ $\dot{N}_{ion}=\dot{N}_{rec}$ at $z=6$
assuming $C_{HII}=10$.

\section{Light from Pop III stars}

Another piece of useful information than can be extracted from the simulation outputs is the relative fraction of normal (Pop II) and massive, metal-free (Pop III) stars.  There are several questions to which we can provide quantitative answer: (i) do some of the current
candidates contain Pop III stars ? (ii) in that case, what fraction of their UV luminosity is powered by them ? (iii) how is this fraction dependent on their $M_{UV}$ luminosity ? (iv) is there a clear observational signature imprinted by Pop III stars ? 

The answer to the first question is straightforward: having analyzed the stellar populations  of the simulated galaxies present at four observationally relevant redshifts, $z=(5, 6, 8, 10)$, we find that a fraction 0.07-0.19 (depending on $z$) of the galaxies contain at least some Pop III stars. We should not emphasize too much on the exact values of this Pop III/Pop II galaxy ratio as fluctuations in galaxy mass and star formation rate might introduce a very large dispersion. The most robust physical quantity to understand the relative importance of the two populations is the ratio of Pop III-to-Pop II star formation rates which is a decreasing function of time (see Fig. 1 of TFS07) never exceeding $10^{-3}$ below $z=10$.

\begin{figure}
\center{\includegraphics[scale=0.42]{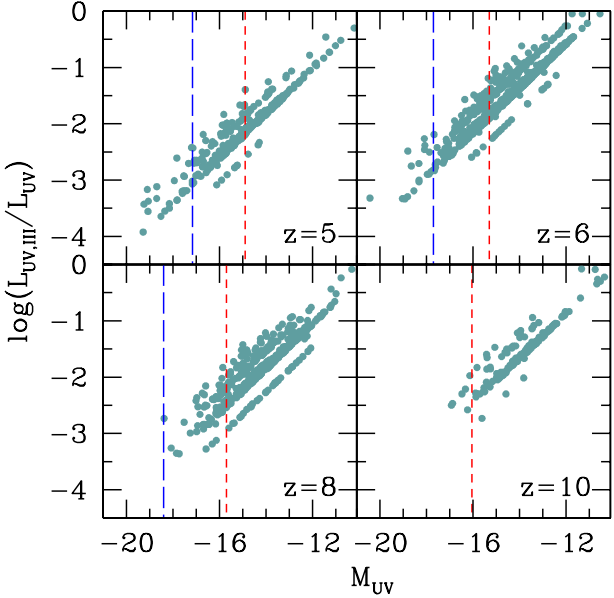}}
\caption{\label{fig:Pop III} Fraction of the total 
luminosity $L_{\rm UV}$ due to Pop III stars $L_{UV,III}$ as a function of the
absolute UV magnitude $M_{UV}$. The vertical short(long)-dashed lines mark 
the sensitivity limit of JWST (HST/WFC3). Absolute UV magnitudes are 
computed as in Fig.~\ref{fig:LF}.} 
\end{figure}

More relevant is question (ii) above, whose answer can be obtained by inspecting Fig. \ref{fig:Pop III},  showing the ratio, $L_{UV,III}/L_{UV}$ of UV luminosities contributed by Pop III and Pop II stars as a function of $M_{UV}$. Light from Pop III stars becomes progressively more important towards fainter objects. This confirms previous findings (e.g. Schneider et al. 2006) that Pop III stars preferentially form in low-$\sigma$ peaks rather than in larger galaxies whose gas has already been enriched by several stellar generations. At the same time, we do not find galaxies containing only Pop III stars: this occurrence is made unlikely by the short lifetimes of such very massive stars. At $z=6$ ($z=10$) the contribution of Pop III stars to the total luminosity is always less than 5\% for $M_{UV}<-17$ ($M_{UV}<-16$). 

Among the candidate galaxies detected so far, we find that Pop III stars contribute less than a few percent to the total galaxy luminosity. Even at the detection limit of JWST, no Pop III-dominated galaxies ($\simgt 50$\% of total luminosity) will be found, due to their extreme faintness ($M_{UV}\simgt -14.$ at $z=6$). However, some objects having $\approx 10$\% of their luminosity powered by Pop IIIs are present at $z=6$ at the 1 nJy sensitivity level reachable by the JWST. 

Recently, many authors (Shapley et al. 2003, Nagao et al. 2008, di Serego Alighieri et al. 2008) have searched for the HeII 1640\AA\, emission line in the spectra of the so-called {\it dual emitters}, i.e. high-$z$ galaxies showing strong emission in both Ly$\alpha$ and HeII lines. The HeII line is usually taken as a well-defined signature of  Pop III stars. Up to now, these searches have given negative results. Motivated by these attempts, we have computed the rest-frame equivalent width of the HeII line for the objects in our simulation box  as
$EW({\rm HeII})=L({\rm HeII})/L_{1460{\rm \AA}}$ where $L(\rm {HeII})=5.67\times 10^{-12} Q({\rm He}^+)$
is the luminosity of the HeII line in erg s$^{-1}$, $Q({\rm He}^+)$ is the rate
of He ionizing photons, and $L_{1460{\rm \AA}}$ is the continuum luminosity at
$\lambda=1460$\AA~ in erg s$^{-1}$ \AA$^{-1}$.
In particular, we focus here on $z=6$ which might be more easily accessible to present or future observations. Regretfully, the perspectives of direct detection of PopIII stars through this technique do not appear as very promising. For objects detectable in the JWST (HUDF) deep field survey, the expected HeII rest-frame equivalent width is $<0.5$ \AA~($<0.1$ \AA). Such small EWs will be very difficult to detect. The EW increases to more accessible values of about $10$ \AA~only if much fainter objects ($M_{UV} = -13$) could be observed. We have to underline that the above discussion implies that we cannot set limits on the total cosmic SFR in Pop III stars using the results of dual emitters searches given that the bulk of Pop III stars may be 'hidden' in galaxies much fainter than any present (and probably future) survey can detect (Scannapieco, Schneider \& Ferrara 2003). 

In conclusion, Pop III stars are found to form essentially at any redshift and
in $ < 20$\% of the galaxies. However, their contribution to the total galaxy 
luminosity is very low, apart from the very faint objects and their detection 
could be extremely difficult even with the next generation of space telescopes.

\section{Early Gamma-Ray Bursts}

Long Gamma-Ray Burst (GRBs) are powerful flashes of $\gamma$-rays that are observed with a frequency of about one per day over the whole sky. The $\gamma-$ray emission is accompanied by a long-lasting tail, called afterglow, usually detected over the whole electromagnetic spectrum. Their extreme brightness easily over-shines the luminosity of their host galaxy and
makes them detectable up to extreme high redshifts, as shown by the discovery of GRB~090423 at $z=8.2$ (Salvaterra et al. 2009b, Tanvir et al. 2009). Metal absorption lines can often be identified in their afterglow spectra, allowing a study of the metal (and dust) content of the environment in which they blow. Finally, once the afterglow has faded, follow-up searches of the GRB host galaxy become possible. At low redshifts,
GRBs are typically found in blue, low-metallicity dwarf galaxies with stellar masses $M_*\sim 10^{8-9}\;\Msun$ and high specific star formation rates (Savaglio et al. 2009). These objects closely resemble the properties of high-$z$ galaxies identified in our simulations, whose 
mean specific star formation rates (SSFR) are $\approx 8-10$ Gyr$^{-1}$, albeit associated with a large spread (for $M_*\ge 10^8$, SSFR=1.5-7 Gyr$^{-1}$ at $z=6-8$). This suggests that 
high-$z$ GRBs can be used as signposts of the same faint galaxies that provide the bulk of the ionization photons in the early Universe\footnote{Note that 
GRBs may be biased tracers of the cosmic star formation rate (e.g. Daigne et al. 2006; Salvaterra \& Chincarini 2007; Salvaterra et al. 2009a). However, if
this is related to the existence of a metallicity threshold for GRB formation of
$Z<0.3\;\Zsun$ as required by the collapsar model (e.g. MacFadyen \& Woosley 
1999) then we expect that high-$z$ GRBs
can be fair cosmic tracers.} 

. Moreover, the study of their afterglows can provide new hints about the metal (and dust) content of the parent galaxy. Finally, it has been proposed that even Pop III stars may eventually blow as powerful GRBs (Fryer, Woosley \& Heger 2001; Yoon, Langer \& Norman 2006; Hirschi 2007; Komissarov \& Barkov 2010). The observation of these Pop III-GRBs may provide a valuable way to detected the elusive, short living first stars. We will discuss these points in the following.

\subsection{GRBs as metallicity probes}
Fig.~\ref{fig:metal} shows the mean stellar metallicity of the simulated sample of 
high-$z$ galaxies as a function of their absolute UV magnitude at different redshifts $z\ge 5$, i.e. the luminosity--metallicity relation for such objects. Although the experimental determination of such a relation would be of the utmost importance in testing theoretical predictions, in practice such effort is hampered by the extreme difficulty to infer $Z$ from lines in the galaxy spectrum with available facilities. 

This problem can be considerably alleviated if a GRB could be found inside of one of these
remote galaxies. In this case absorption features produced by heavy elements dispersed in the interstellar medium (ISM) surrounding the GRB would leave a characteristic and recognizable imprint on top of the GRB afterglow spectrum. Moreover, there is now a good agreement on that GRBs are usually hosted in relatively small, star-forming galaxies (Savaglio et al. 2009): these requirements would make the dwarf high-$z$ galaxies which, according to our findings, are the dominant population during the first cosmic billion year, optimal candidate hosts. 

The only high-quality spectra burst at $z>6$ available so far, GBR~050904 at $z=6.3$ (Kawai et al. 2006), shows the expected metal absorption features, witnessing the presence of metals at the same redshift of the GRBs (Kawai et al. 2006). From this data, and with the further assumption that the measured sulfur ([S/H]=$-1.3\pm 0.3$) is a good proxy for metallicity, one can determine the metal content of the host galaxy. Berger et al. (2007) attempted a detection of the GRB host galaxy with HST and {\it Spitzer}, and were able to set an upper limit to the host luminosity of 
$M_{UV}\simgt -20.3$, i.e. $M_{UV}\simgt M_{UV}^*$ at $z=6$. By combining the HST and {\it Spitzer} upper limits, they set an upper limit to the stellar mass of the galaxy, $M_*\simlt {\rm few}\times 10^9\;\Msun$.  This constraint is consistent with the predictions presented in Fig. \ref{fig:metal} (notice the data point corresponding to the Berger et al. 2007 determination), which might indicate that at $z\approx 6$ a galaxy with $Z\approx 0.05 Z_\odot$ is on average 2-3 magnitudes fainter than the detection limit of the experiment; in addition, such metal abundance level is quite typical for galaxies located at that epoch. 

GRB~090423 (Salvaterra et al. 2009b, Tanvir et al. 2009) has been recently detected at $z=8.2$. The intrinsic properties of the burst (both of the prompt and afterglow phases) are similar to those observed at low/intermediate redshifts, suggesting that the progenitor and the medium in which the burst occurred are not markedly different from those of low-$z$ GRBs 
(Salvaterra et al. 2009b, Chandra et al. 2010). Chary et al. (2009) observed the field of GRB~090423 for 72 hours using {\it Spitzer}/IRAC at 3.6 $\mu$m, looking for its host galaxy. 
The observation was performed $\sim 46$ days after the GRB trigger, (corresponding to about 5 days in the burst rest-frame). A weak source was detected at the location of the GRB
afterglow with  $L_{AB}=27.2\pm 0.3$. The observed flux is consistent with the power-law decay of the GRB afterglow, suggesting that the source is still contaminated by the GRB emission and implying a limit on the absolute magnitude of the host of $M_{3900{\rm \AA}}>-19.96$. 
From Fig.~3, we expect that the host galaxy of GRB~090423 should be enriched at a level of a few percent solar. This relatively high metallicity may explain the high equivalent neutral hydrogen column density measured in the X-ray afterglow, although a wide range of systematic effects or the contamination by low-$z$ intervening absorption systems may be a more valuable, 
alternative explanation for the observed absorption (Chandra et al. 2010).

These examples nicely show how GRBs represent an unique tool to study the high-$z$ Universe. In particular, the observation of their optical-NIR afterglow may allow to eventually study the evolution of the mass-metallicity relation up to very high-$z$. To achieve this goal, 
high resolution and good signal-to-noise afterglow spectra are required as soon as possible 
after the GRB detection in $\gamma$-rays. As nicely demonstrated by GRB~090423, GRBs are 
easily detectable well beyond any other astrophysical object. Indeed, up to $\sim 5$\% of 
all GRBs detected by the {\it Swift} satellite are expected to be at $z>6$ (Salvaterra \& Chincarini 2007, Salvaterra et al. 2009b). Future missions (e.g. EXIST, XENIA, SVOM) will rapidly increase the high-$z$ GRB sample (Salvaterra et al. 2008) allowing a statistical study of GRB hosts in the high-$z$ Universe and a direct check of the galaxy metal enrichment history at those early epochs. In particular, EXIST, thanks to its 1.1m optical-NIR 
telescope, will be able to take the GRB afterglow spectrum only 300 s after the trigger, allowing an on-board direct measure of the redshift and the identification of metal  absorption lines when the afterglow is still sufficiently bright even for high-$z$ bursts (Grindlay et al. 2009).

\subsection{GRBs from Pop III stars}

It has been widely discussed (see Bromm \& Loeb 2007 for a review) whether GRBs can arise from the collapse of  massive, metal-free stars. While the large envelopes of these objects may suppress the emergence of relativistic jets out of their surface, the weak winds expected for 
low-metallicity stars can prevent angular momentum loss during their evolution, producing the rapidly rotating central configurations needed to produce a GRB. In the following we will assume that massive Pop III stars produce GRBs.

Excluding Pop III progenitors in the mass range [140,260] $\Msun$, that will explode as Pair Instability Supernovae (PISN, Heger \& Woosley 2002) leaving no remnant, we have two possible channels through which GRBs can occur: (i) 100-140 $\Msun$ (Yoon et al. 2006, Hirschi 2007) and (ii) 260-500 $\Msun$ (Fryer et al. 2001, Komissarov \& Barkov 2010). In the latter case, since both the luminosity and the duration are thought to be proportional to the black hole mass, an extremely bright and long GRB is expected (Fryer et al. 2001), while more typical luminosities and durations may be expected in the case of smaller progenitors (Hirschi 2007).

We can estimate a strong upper limit for the rate of Pop III-GRB detections as follows. Suppose a fraction $f_{GRB}$ of all Pop III stars with masses in the ranges discussed above produce GRBs with a typical beaming angle $\theta$, and that all Pop III-GRBs are detectable by present-day satellites given the their extreme brightness. Then the observed rate of Pop III-GRB, $R_{GRB}(>z)$, is given by
\begin{equation}
R_{GRB}(>z)=5.5\times 10^{-3} \left(\frac{\theta}{6^\circ}\right)^2 \int_z dz\, f_{GRB}\, \eta\,
\frac{\dot{\rho}^{III}(z^\prime)}{1+z^\prime} \frac{dV}{dz}
\end{equation}
where $\dot{\rho}^{III}(z)$ is the total Pop III star formation rate at redshift $z$, $\eta=\int_{M_{min}}^{M_{max}} \phi(m_*) dm_*/\int_{100}^{500} m_* \phi(m_*) dm_*$ is the number of Pop III stars with masses in the range $[M_{min},M_{max}]$ given the assumed IMF $\phi$, and $V$ is the cosmological volume per unit solid angle. The factor $(1+z)^{-1}$ accounts for time dilation effects due to redshift. The computed rate of Pop III-GRB at $z>6$ is then 
$R_{GRB}\sim 7.5 (2.5) f_{GRB}  (\theta/6^\circ)^2$ yr$^{-1}$ sr$^{-1}$ for GRB progenitors in the mass range 100-140 $\Msun$ (260-500 $\Msun$). This rate is of the same order of that expected for normal (i.e. Pop II/PopI progenitors) GRBs (Salvaterra \& Chincarini 2007, Salvaterra et al. 2009b). However, we stress here that very likely $f_{GRB}\ll 1$, resulting in lower detection rates. The detection of one of such Pop III-GRBs might represent the most 
promising way to directly detect the very first stars to have formed.

\section{Critical discussion}
The present results allow to build a coherent and quantitative description of the properties of elusive high-redshift, possibly primordial, galaxies. They are also very useful to interpret the data coming from deep surveys as the HST/WFC3 and future ones. However, there is considerable room for improvement left by our study. In the following we would like to elaborate on the uncertainties and shortcomings of our findings.

We first note that resolving the dwarf galaxy population and following the PopIII transition process along with the large variety of physical processes implemented in the simulation limits the size of the cosmic volume that can be simulated. Resolution is certainly an important issue, as it is well known (see, e.g. Governato et al. 2010) to affect the simulated star formation rates and cause the loss of sub-galactic structures. The dependence of the results from resolution has been presented an analyzed for the same set-up of the present simulations in Tornatore et al. (2007) to which we refer the interested reader (see Fig. 1 of that paper). Resolution might also alter the details of the "Pop III wave" evolution, since galactic substructure allows star formation to occur at the edges of galaxies as well. 
The PopIII-PopII transition is also very dependent on the assumed IMF
 of PopIII stars. Our conclusions are valid under the assumption that
 PopIII stars were very massive ($M\ge 100\;\Msun$) and the first metal
 production is driven by the explosion of pair-instability SNe
 (see Schneider et al. 2006 for alternatives).

Our box is also too small to properly describe cosmic reionization, let alone that we are not even attempting to properly treat radiative transfer.  These issues have been already addressed in previous works of our group; as already stated we are concerned here with the properties of high-z galaxies, which are presumably more affected by their internal physics rather than by the environment, as we explain below. Comparing the simulated volumes with the observed ones is very challenging as considerable uncertainty exists on the latter (see discussion in Appendix B of Bouwens et al. 2009). However, as it could be induced from the extension of the LF towards the most luminous and rarest objects at z=7, we estimate that the volume sampled by experiments should be about 30 times larger than our simulated one.

\begin{figure}
\center{\includegraphics[scale=0.42]{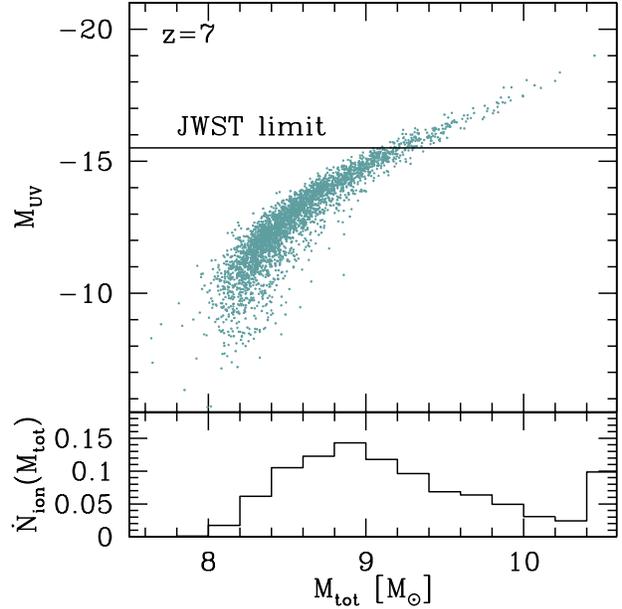}}
\caption{\label{fig:mhalo} {\it Upper panel:} $M_{UV}$ vs. total (dark+baryonic) mass relation for simulated galaxies at $z=7$; the JWST detection limit is shown. {\it Lower panel:} differential fractional ionizing photon rate distribution as a function of total galaxy mass. The last bin on the left shows the contribution of larger halos too rare
to be caught in our relatively small simulation box, computed by 
integrating the observed LF down to $M_{UV}=-25$.}
\end{figure}

The next caveat comes from the fact that we have neglected the effects of minihalos (virial temperature $< 10^4$ K). Our resolution does not allow us to track the formation of such objects, whose stellar contribution remains very uncertain due to radiative feedback effects (Haiman \& Bryan 2006; Susa \& Umemura 2006; Ahn \& Shapiro 2007, Okamoto, Gao \& Theuns 2008, Salvadori \& Ferrara 2009). The presence of such small collapsed structures, if able to form stars, could alter the reionization history to some extent and increase the number counts of high-redshift galaxies, if detectable. As far as reionization is concerned, it has already been shown by Choudhury \& Ferrara (2007) that acceptable fits to all relevant reionization data can be obtained without any need for PopIII stars. The bulk of the ionizing photons in those models is produced by halos with virial temperatures just above $10^4$ K, with increasingly better solution if normal, PopII stars are allowed to form in minihalos. As shown in Fig. \ref{fig:mhalo}, where we plot the UV magnitude as function of the total mass of z=7 galaxies, sources detectable with JWST have halo masses $M_h \simgt 10^9 \Msun$. This corresponds to circular velocities $v_c \simgt 50$ km s$^{-1}$. These objects are large enough that suppression by
UVB photoionization filtering is at best marginal, if not negligible at all,
as most of the works above agree upon. Internal mechanical feedback might be indeed more important. This process is however is already included at best in the simulations when computing the star formation rate of individual galaxies, modulo the many uncertainties that still plague our understanding of such phenomenon.  

\section{Summary}

By using high resolution simulations specifically crafted to include the relevant physics of galaxy formation, along with a novel treatment of the metal dispersion that allows us to follow the PopIII-PopII transition as dictated by the critical metallicity scenario, we have been able to reproduce the observed UV LFs over a wide redshift range, $5 < z < 10$. We have also shown, by combining the simulation outputs with a dust model previously developed for LAEs (Dayal et al. 2010),
that dust effects at $z\approx 7$ should be marginal, although this statement depends on many details (the dust properties and the dust distribution scale in the ISM) that are poorly constrained at this time.    
The general picture that can be drawn from our investigation is broadly consistent with the available data and therefore can be used to make specific predictions for the JWST. 
It is then useful to schematically summarize the main findings of the present work:

\begin{itemize}
\item{The simulated high-$z$ galaxy UV LFs match remarkably well with the amplitude and
slope of the observed LFs in the redshift range $5 < z < 10$.}
\item{The LF shifts towards fainter luminosities with increasing
redshift, mimicking a pure luminosity (or density) evolution. The faint-end
slope of the LF does not vary from $z=5$ to $z=10$, keeping an almost constant
slope value of $\alpha=-2$.}
\item{Many galaxies at $z\approx 7$, especially the smallest ones, are virtually dust-free, and none of them shows dust extinctions larger than  $E(B-V)=0.009$. This evidence allows us 
to safely neglect the effects of dust on the UV LF.}
\item{The stellar population of high-$z$ galaxies shows typical ages in the range 
100-300 Myr at $z=5$ and 40-130 Myr at $z=7-8$, implying that they started to form 
stars as early as $z=9.4$. These objects are enriched rapidly with metals and galaxies identified by HST/WFC3 show metallicities $\approx 1/10 \Zsun$ even at $z=7-8$. 
The trend of decreasing metallicity (and increasing spread) towards low mass halos indicates that small galaxies are more affected by supernova feedback and loose a larger fraction of the heavy elements they produce.}
\item{The relation between the star formation rate and stellar mass of simulated follows an almost linear relation with significant spread towards the lowest masses, implying an almost constant specific star formation rate.}
\item{The bulk of the ionizing photons is produced by objects populating the 
faint-end of the LF. These galaxies are beyond the capabilities of current
survey, but JWST will be able to resolve them up to $z=7.3$.}
\item{Massive PopIII stars continue to form essentially at all redshifts and
in $<20$\% of the galaxies. However, their contribution to the total galaxy 
luminosity is negligible ($<$ 5\%) for all objects with the marginal exception of the 
extremely faint ones; their detection will be tremendously difficult even for
the next generation of space telescopes.}
\item{The typical high-$z$ galaxies closely resemble the GRB host population observed 
at lower redshifts. This fact suggests that GRBs can be used to detect and study these objects, providing unique information about the first stages of structure formation. In particular, they can be used to extend the study of the mass-metallicity and its evolution 
to very high redshifts. Moreover, if PopIII stars end their lives in a GRB explosion, the
detection of one of these bursts might represent the most promising way to directly detect 
the very first stars.}
\end{itemize}

\section{Acknowledgments}
We thank R. Schneider and L. Tornatore for collaborative support. Discussions and the stimulating environment at {\tt DAVID IV}, held at OAArcetri, Florence is kindly acknowledged.

\end{document}